\documentclass[12pt]{article}
\usepackage{graphicx,amsmath,amssymb,cite}
\usepackage{epsfig,epsf}
\usepackage{graphicx}
\usepackage{epstopdf}
\newcommand{\beq}{\begin{equation}}
\newcommand{\eeq}{\end{equation}}

\numberwithin{equation}{section}

\begin{document}

\begin{flushright}
DESY 11-149\\
SHEP-11-24\\
\end{flushright}

\vspace*{0.5 cm}

\begin{center}

{\Large{\bf Indirect Evidence for New Physics
 at the 10 TeV Scale }}

\vspace*{1 cm}

{\large H. Kowalski~$^1$, L.N. Lipatov~$^{2}$, and D.A. Ross~$^3$} \\ [0.5cm]
{\it $^1$ Deutsches Elektronen-Synchrotron DESY, D-22607 Hamburg, Germany}\\[0.1cm]
{\it $^2$ Petersburg Nuclear Physics Institute, Gatchina 188300, St. Petersburg, Russia}\\[0.1cm]
{\it $^3$ School of Physics and Astronomy, University of Southampton,\\Highfield, Southampton SO17 1BJ, UK}\\[0.1cm]
 \end{center}

\vspace*{3 cm}

\begin{center}
{\bf Abstract} \end{center}

We show that the supersymmetric extension of the Standard Model modifies the structure of the low lying BFKL discrete pomeron  states (DPS) which give a sizable contribution to the gluon structure function in the HERA $x$ and $ Q^2$ region. The comparison  of the gluon density,  determined within DPS with N=1 SUSY, with data  favours a supersymmetry scale  of the order of 10 TeV.      The DPS method described here could open a new window to the physics beyond the Standard Model. 

\vspace*{3 cm}

\begin{flushleft}
  September 2011 \\
\end{flushleft}

\newpage

\section{Introduction}
In our previous paper \cite{KLRW} we have shown that HERA $F_2$ data, at low $x$, can be very well described by the gluon density constructed from the discrete spectrum of eigenfunctions of the BFKL kernel, i.e. from the pomeron wave functions. This  first successful confrontation of the BFKL formalism \cite{BFKL} with data led to the
 unexpected question as to whether the HERA data are sensitive to the Beyond Standard Model (BSM) effects. These effects, although only present at scales  that are much higher than the  region of HERA  data, can nevertheless affect the quality of the fits to data because the shape of many of the contributing   eigenfunctions   has an apparent sensitivity to the BSM effects. This apparent sensitivity is due to the fact that the support of  eigenfunctions extends to very high transverse momenta where BSM effects have to be present. Since the eigenfunctions are constructed in a global way,
i.e. the behaviour of the eigenfunctions at energies way above the threshold feeds into their behaviour
at  low energies,  these eigenfunctions will be sensitive to any BSM physics.

In this paper we investigate whether this possible sensitivity to BSM effects indeed exist, as it
  also depends on an  adequate treatment of the infrared boundary condition. As a popular example of BSM effects we have chosen the N=1 supersymmetry.  For this purpose we modified the beta function and the kernel  of the BFKL equation to include the contributions from the  superpartners and confronted the modified gluon density   with data.

The paper is organized as follows; in Section 2 we give a brief summary of the properties of the discrete
pomeron solution to the BFKL equation. In Section 3 we describe the construction of the infrared boundary condition and the changes introduced by the two-loop running of the coupling. In Section 4 we discuss  the effects of the supersymmetric changes of the $\beta$-function and of the  eigenfunctions of the BFKL kernel. In Section 5 we present and discuss the results and in Section 6 we describe the properties of the determined infrared boundary. In Section 7 we give a summary.

\section{The Discrete Pomeron Solution to the BFKL Equation}
In this section we give a brief summary of the properties of the discrete
pomeron solution to the BFKL equation,  described in detail in
\cite{KLRW}.

The BFKL amplitude for 
the scattering of high-energy gluons with transverse momenta $\mathbf{k}$
and $\mathbf{k}^\prime$,
 is a Green function constructed from the discrete eigenfunctions of the BFKL  kernel,
i.e. the solutions $f_n(k)$ to the equation
 \beq  \int d^2\mathbf{k}^\prime 
{\cal K}\left(\mathbf{k},\mathbf{k}^\prime, \alpha_s(k^2)\right)
 f_n(\mathbf{k}^\prime) \ = \ \omega_n  f_n(\mathbf{k}^\prime). \label{ev}  \eeq
 where $\alpha_s(k^2)$
is the strong coupling which runs with the magnitude, $k$,
of the transverse momentum of one of the gluons.
 This running leads to
an oscillatory eigenfunction, $f_n$, whose frequency, $\nu_n$, in the semiclassical approximation, depends on
the transverse momentum, $k$, so that the eigenvalues, $\omega_n$, are given in
terms of the LO and  NLO characteristic functions of the oscillation frequency $\nu$,
  \beq 
  \omega \ = \  \left(\frac{\alpha_s(k^2) C_A}{\pi} \right)\chi_0(\nu) +
    \left( \frac{\alpha_s(k^2) C_A}{\pi} \right)^2 \chi_1(\nu) + \cdots
\label{evs},\eeq
where for the moment we have ignored the resummation of collinear divergences in the NLO
characteristic function \cite{salam}.
The frequency  depends on $\omega$ and
 decreases as $ k$ increases, reaching a critical
point $k_{crit}$ at $\nu(k_{crit})=0$ where it changes from real to imaginary values. Below $k_{crit}$ the  eigenfunction has an oscillatory behaviour but above $k_{crit}$
it  decreases exponentially with $\ln k$.
 The matching of the phases immediately below and above the critical point  fixes the
phase of the oscillations at $k_{crit}$ to be $-\pi/4$ (this being the phase of the
Airy function, which was shown in ref.\cite{KLRW,EKR, lipatov86}
to provide a very good approximation to the eigenfunctions). This 
 phase, $\phi(k)$,
at any lower value of $k$ is then determined by integrating the $k$-dependent
frequency from $k_{crit}$ to $k$, namely
 \beq \phi(k) \ = \ -\frac{\pi}{4} + 2 \int_{k}^{k_{crit}} \nu(k) d\ln(k) . \label{phase} \eeq 


For a given value of $\omega$ the phase, $\eta \, \pi$, of the oscillation at some
 infrared transverse momentum, $k_0$,
is determined from the perturbative  BFKL dynamics (with running coupling), through eqs.(\ref{ev},\ref{evs}).  We make a very general assumption that the infrared (non-perturbative) properties of QCD impose some phase at $k_0$, defined up to an ambiguity of $n\pi$,
which can also depend on $\omega$. We find then that we can only match this phase to that determined from eq.(\ref{phase}) for one value 
 of $\omega$, for each 
 integer $n$, where $n$ corresponds to the number of 
 turning points of the eigenfunctions. This leads to the quantization of the
 spectrum (i.e. discrete pomeron poles), in  keeping with the predictions of Regge theory.

Before a comparison can be made with the measured structure function, $F_2$,
 it is necessary to convolute this
Green function with the impact factors for the virtual photon at one end and for
the proton at the other, (see Section 6 of ref.\cite{KLRW}). The impact factor for the virtual photon is
calculable in perturbative QCD and has support, which is peaked at transverse
momenta of the order of the $Q^2$ argument of the structure function,
whereas the proton impact factor cannot be so calculated and is
assumed to have a simple form with support up to ${\cal O}(1)$  GeV.

One of the main results of ref. \cite{KLRW} was
that a  very good quality fit to HERA-$F_2$ data \cite{H1ZEUS}
 (with $Q^2$ above $8 \, \mathrm{GeV}^2$)
is obtained  taking a very simple form for the dependence of the above-mentioned (non-perturbative) infrared phase,
$\eta_n$, on eigenfunction index $n$. It was found that in order to obtain
this good description of data, it was necessary to take around 120 eigenfunctions
of the BFKL kernel. 


Although the oscillation frequency varies with transverse momentum, $k$, the period, $\Delta$,
of oscillation  (in $\ln(k)$) defined by
 \beq 2 \int_{\ln(k)}^{\ln(k) + \Delta} \nu(k^\prime) d\ln(k^\prime) \ = \ 2 \pi \eeq
turns out to be roughly constant ($\Delta \, \sim \, 8$), beyond the first two
turning points. This means that the main difference between the $n^{th}$ and $(n+1)^{th}$
(for $n \, > \, 2$) eigenfunction is that the latter has one more half period, which leads to
a rapid increase in the critical momentum, $k_{crit}$ with eigenfunction index $n$
 \beq k_{crit} \ \sim \ c\cdot e^{4n}\label{range},\eeq 
 where $c$ is a constant of the order of $\Lambda_{QCD} $.
For the first eigenfunction the value of $k_{crit}$ is ${\cal O}( 10 \, \mathrm{GeV})$.
It therefore follows that $k_{crit}$ rapidly exceeds the threshold for most postulated theories beyond the Standard
 Model. On the other hand, if a threshold for new physics does indeed exist, the oscillation
frequency is affected above this threshold and consequently the oscillation phase at {\it all} lower
transverse momenta will be altered thereby affecting the matching of the phase to the phase
imposed by the infrared dynamics of QCD. This in turn modifies the  pomeron spectrum, $\omega_n$.
 It is in this sense that a modification of the
high-energy  behaviour of the eigenfunctions ``feeds'' into the low-energy behaviour.


This immediately posed the question as to what the
effects would be on the quality of the fit, if there were some new physics
 far above the energy scale  of HERA.
 

\begin{figure}
\centerline{\epsfig{file=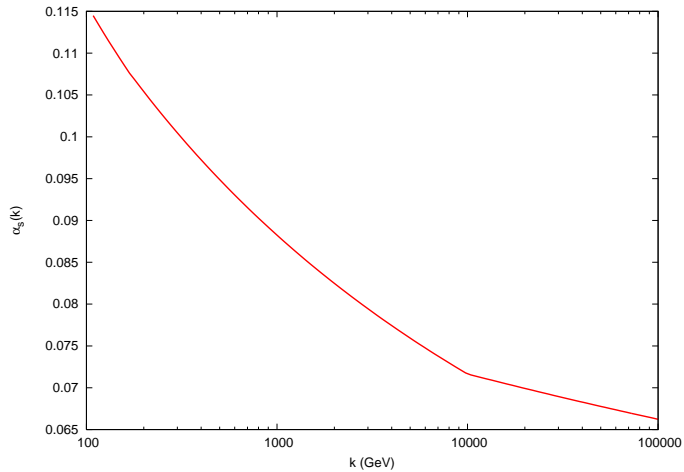, angle=270, width = 9 cm}}
\caption{ The running of $\alpha_s$ across a threshold for N=1 SUSY at 10 TeV \label{alphahigh} }
\end{figure}

\section{The Infrared Boundary}

In ref.\cite{KLRW}, we defined the infrared boundary as a phase condition at the lowest possible  value of the transverse momentum, $k=k_0$, which can be safely reached by the perturbative calculation. To make this value as close as possible  to $\Lambda_{QCD}$ we  considered only the one-loop running of the coupling. This gave a value of $k_0=0.3$ GeV, which corresponds to $\alpha_s \sim 0.7$. The reason for running the coupling at one-loop only  was that in principle this is the same order of perturbation
theory as the NLO characteristic function, $\chi_1$ \cite{FL}.

However, given that we modify eq.(\ref{ev}) by resumming all the large
corrections in $\chi_1$ using the technique of ref.\cite{salam}, 
it is more appropriate to take the $\beta$-function  to two-loop order
which is what we use in this paper.

When we do this, we are faced with a problem - namely that we cannot run the coupling
below an ``infrared'' scale $k_0=0.6$ GeV, which corresponds to $\alpha_s\sim 0.7$ 
(at the two loop level), without approaching the Landau pole too closely.
On the other hand, the infrared boundary conditions are to be imposed at a transverse momentum of order $\Lambda_{QCD}$.
 Moreover we need to know  the eigenfunctions below $k_0$ in order to
perform a convolution with the proton impact factor, which has support mainly below
the $k_0$ value. Therefore, guided by the behaviour of the eigenfunctions in  perturbative QCD,
 we continue them  down to a lower momentum $\tilde{k}_0$, which should be of order $\Lambda_{QCD}$,
  using the extrapolation
of the phase  $\phi_n(k)$ 
\beq \phi_n(\tilde{k}_0) \ = \ \phi_n(k_0) - 2 \nu_n^0 
\ln\left(\frac{k_0}{\tilde{k}_0}\right),
\label{eq:phrel}
 \eeq
where for each eigenfunction, with index  $n$, $\nu_n^0$ is the frequency of the oscillations
near $k=k_0$~\cite{KLRW}. We have assumed that this frequency is constant below $k_0$, an assumption
which is correct for sufficiently small $k_0$,
at least for the leading order BFKL kernel (see \cite{lipatov86}). 
Any deviation from constant frequency should have a negligible effect as we are only extrapolating over a small range in gluon transverse momentum. 
The numerical values of $\nu_n^0$ 
are obtained by inverting the eigenvalue equation (\ref{evs}), modified according to \cite{salam}. 

\section{N=1 Supersymmetry at various Thresholds}
The ``new physics'' that we investigate in this paper is the popular $N=1$ supersymmetric extension
of the Standard model above a given threshold $k_T$, which for simplicity we assume to be
a common mass threshold for all  superpartners.
 Below this threshold the running of the
coupling is governed by the $\beta$-function to two-loop order
\beq \beta_< \ = \ -\frac{\alpha_s^2}{4\pi} \left(\frac{11C_A}{3}-\frac{2}{3} n_f \right)
 - \frac{\alpha_s^3}{(4\pi)^2} \left(\frac{34 C_A^2}{3}+\left(\frac{10C_A}{3}+2C_F\right) n_f\right),
\eeq
where for the case of QCD, $C_A=3, \ C_F=4/3$ and $n_f$ is the number of active flavours.
Above the threshold, the beta function is given by
\beq \beta_> \ = \ -\frac{\alpha_s^2}{4\pi} \left(3C_A -  n_f \right)
 - \frac{\alpha_s^3}{(4\pi)^2} \left(6 C_A^2+\left(-\frac{2C_A}{3}+2C_F\right) n_f\right).
\eeq
This leads to a ``kink'' (discontinuity in the derivative) in the running of $\alpha_s$
at the threshold for N=1 SUSY, which can be seen in Fig.\ref{alphahigh}.

\begin{figure}
\epsfig{file=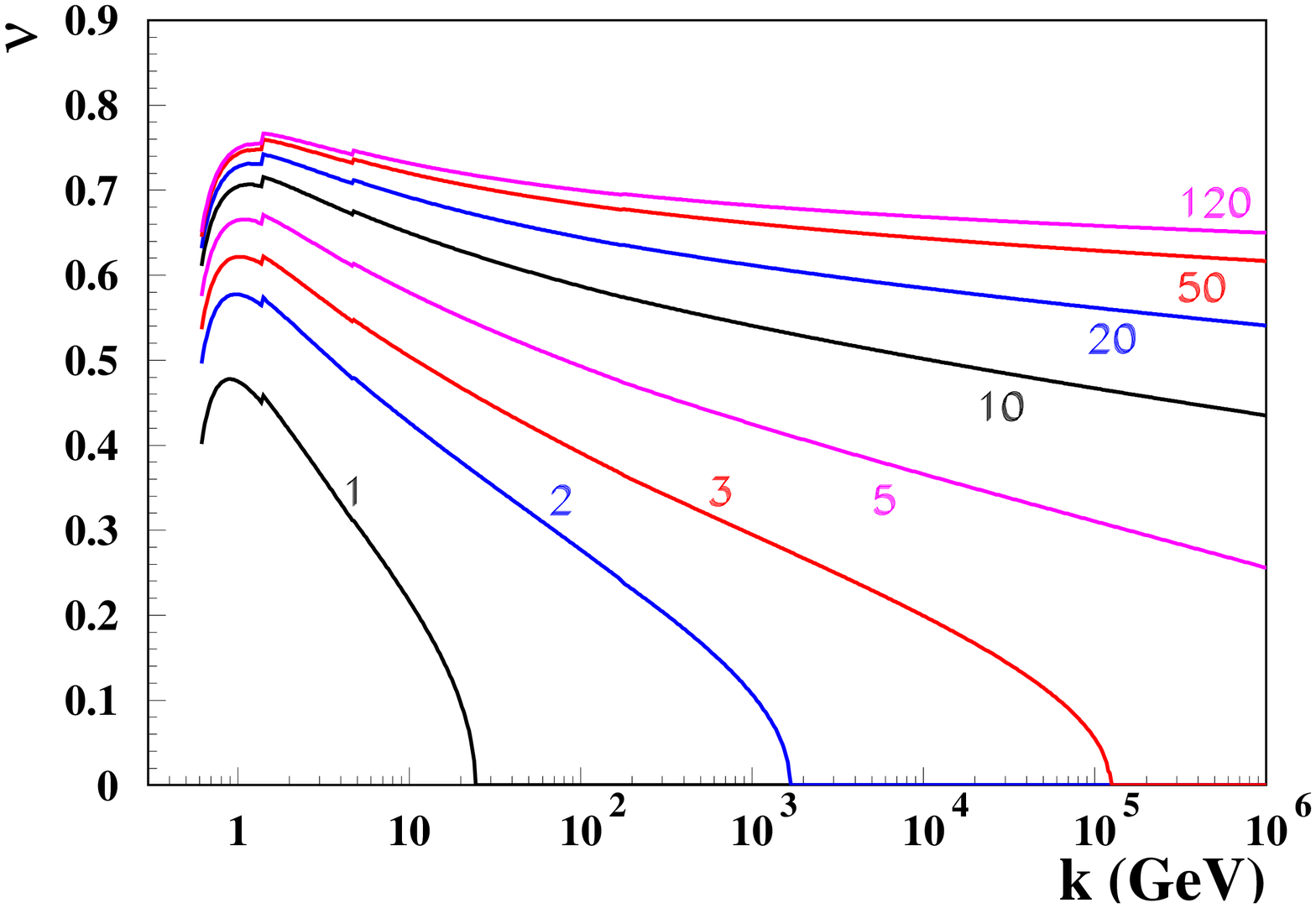, width = 8.5 cm} \
 \epsfig{file=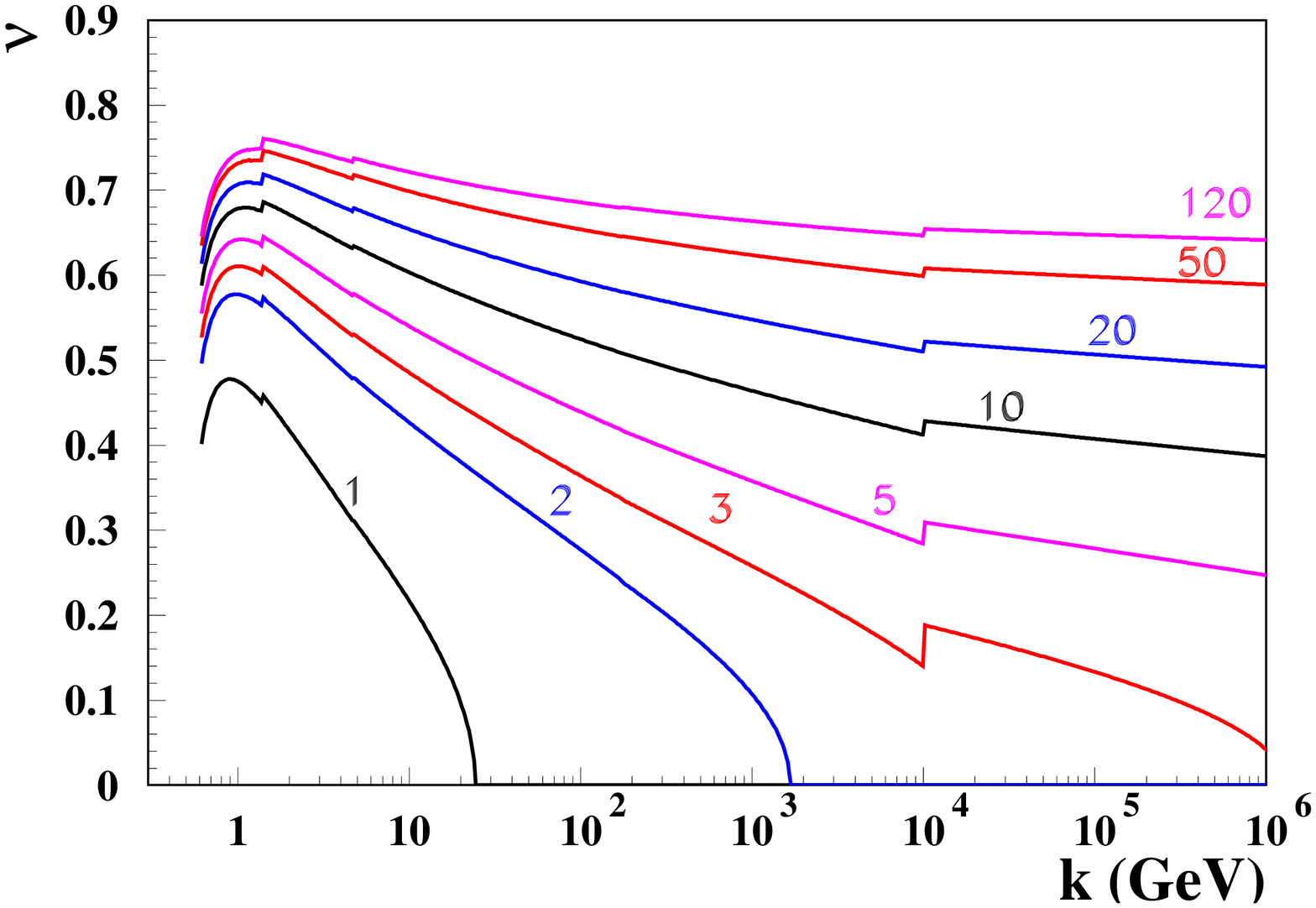,  width = 8.5 cm}
\caption{Oscillation frequencies as a function of gluon transverse
 momentum for various eigenfunctions. The left-hand pane is the case of the Standard
Model and the right-hand pane is the case of N=1 SUSY above a threshold of 10 TeV. For 
the purpose of this comparison it has been assumed that the infrared phases are the same
in both cases.
  \label{nufig}}
\end{figure}

The fact that the coupling runs more slowly above the SUSY threshold means
that the oscillation frequency varies more slowly with $k$ and therefore
 the critical
transverse momentum, $k_{crit}$ (where $\nu=0$), is pushed out  further away. Thus,
for example,
 in the case
of a SUSY threshold at 10 TeV, if we assume that the phase of the oscillations is zero
at $k_0=0.6$ GeV~\footnote{This phase is merely an  example designed to illustrate the
difference in the behaviour of the eigenfunctions for the Standard model and the N=1 SUSY model.
In practice these phases are determined from the fit to data and are different for the two
 models.},  
the first two eigenfunctions are identical as the critical
momentum is below the threshold, whereas for the third eigenfunction the critical momentum
is at $k_{crit}=1.2 \times 10^5 \, GeV$ in the case of the SM but at 
$k_{crit}=1.3 \times 10^6 \, GeV$ in the case of SUSY. This can be seen from Fig.\ref{nufig}.

Furthermore the NLO characteristic function, $\chi_1(\nu)$ acquires an additional contribution
\cite{kotlip} of
\beq \delta_f \chi_1(\nu) \ = \ \frac{\pi^2}{32} 
  \frac{\sinh(\pi \nu)}{\nu (1+\nu^2) \cosh^2(\pi \nu)}
   \left(  \frac{11}{4}+3\nu^2  \right) \eeq
from the octet of Majorana fermions  (gluinos), and
\beq \delta_s \chi_1(\nu) \ = \ -\frac{\pi^2}{32} \frac{n_f}{C_A^3}
  \frac{\sinh(\pi \nu)}{\nu (1+\nu^2) \cosh^2(\pi \nu)}
   \left(  \frac{5}{4}+\nu^2  \right) \eeq
from the squarks. Note that it is this discontinuity in $\chi_1$ at the SUSY threshold which is
responsible for the discontinuities in  the frequencies at threshold and {\it not}
the change in the rate of running of the coupling, which  remains a continuous 
function\footnote{ A similar smaller discontinuity can be seen at around 3 GeV. This corresponds
to the c-quark threshold. There are analogous, even smaller, discontinuities
  at the b-quark and t-quarks thresholds}.
The change in frequency thus compensates for the change in the characteristic
function in order to ensure that the eigenvalues, $\omega_n$ remain unchanged as one passes
through the threshold \footnote{The discontinuous changes in frequency are due to the fact that
the change in characteristic function is imposed at a threshold in its entirety. A determination
 of the NLO characteristic function which accounted for the mass of internal particles
would smooth out these discontinuities.}.

\begin{figure}
\centerline{\epsfig{file=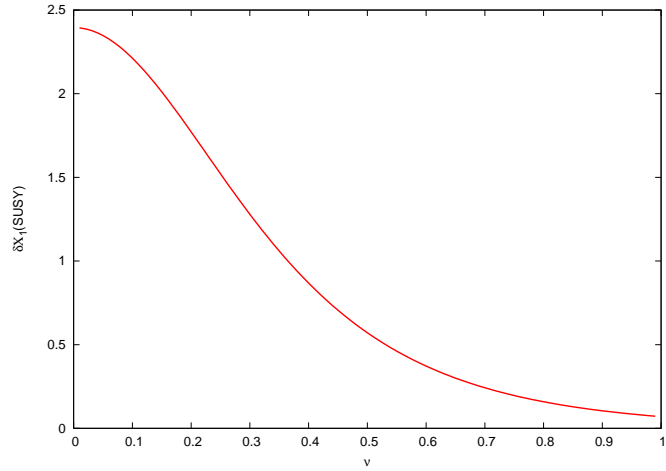, angle=270, width=9cm}}
\caption{ Increase in the NLO characteristic function, $\chi_1$ as
a function of frequency $\nu$ \label{dchi1}}
\end{figure}

The contribution, $\delta \chi_1$,  of these additional terms to $\chi_1$
 is shown as a function of frequency in Fig. \ref{dchi1}
where it can be seen that this is a rapidly decreasing function, which explains why the
discontinuities in frequency 
 at threshold are much larger for the lower eigenfunctions for which the frequency at threshold is 
lower.

\begin{figure}
\centerline{\epsfig{file=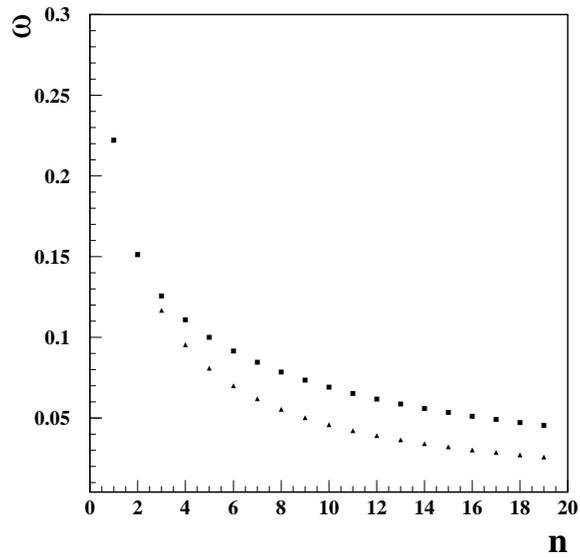, width=9cm}}
\caption{The first 20 eigenvalues in the case of the Standard model (triangles)
 and SUSY at a threshold of 10 TeV (squares) \label{omegan}}
\end{figure}

The differences in frequencies also affect the magnitude of the eigenvalues, beyond the first two,
as can be seen in Fig. \ref{omegan}.
The simplest way to understand this is to consider the value of $\alpha_s$ at
$k_{crit}$
 for the two models. Although $k_{crit}$ is an order of magnitude larger for the SUSY model,
the fact that the coupling runs more slowly actually means that the value of the coupling
at $k_{crit}$ is slightly {\it larger} for the SUSY model. Moreover the  NLO characteristic function
is larger in the SUSY model. These two effects combine to produce somewhat larger eigenvalues -
in the case of the third eigenvalue the difference is about 0.01. For the higher eigenfunctions,
for which $k_{crit}$ is sufficiently large,  eq.(\ref{evs}) is a valid approximation
(at $k_{crit}$)
without collinear resummation,
we may approximate the difference, $\delta\omega_{12}$,
 in eigenvalue between the two models
 in terms of the difference of the running couplings, $\alpha_s(k_{crit \, 1})-\alpha_s(k_{crit \, 2})$,
 at the value of $k_{crit}$ for each model (where the frequency vanishes), namely 
 \beq \delta \omega_{12} \ \approx \ \left( \alpha_s(k_{crit \, 1}) - \alpha_s(k_{crit \, 2}) \right)
 \frac{C_A}{\pi}\chi_0(0) +  \left( \frac{ \alpha_s(k_{crit \, 1}) C_A}{\pi}\right)^2 \delta \chi_1(0),
 \eeq
which gives numerical results in agreement with those seen in Fig. \ref{omegan} for $n \, > \, 10$.

\begin{figure}
\centerline{\epsfig{file=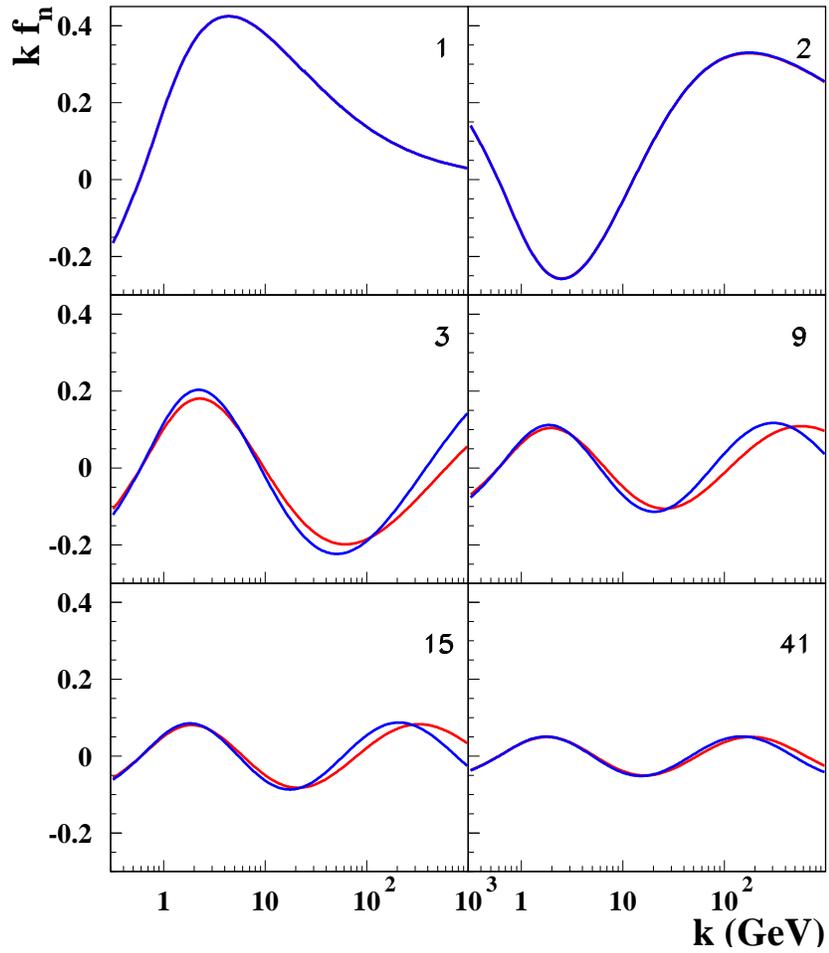, width=12cm}}
\caption{Comparison of  a representative subset of eigenfunctions
 in the Standard Model (blue) and the SUSY model (red) computed with the same value of $\eta =0$ . The eigenvalue number is given in the upper right corner.  \label{phiall}}
\end{figure}

In Fig. \ref{phiall} we show a representative subset of  eigenfunctions in the Standard Model  and the SUSY model in the transverse momentum region relevant for a fit to HERA data. 
The eigenfunctions are computed at the same value of $\eta=0$ to display SUSY effects only (in the fit the eigenfunctions with the same $n$ have in general somewhat different $\eta$'s).
As expected, the first two
eigenfunctions are identical since their values of $k_{crit}$
are below the SUSY threshold. The third and higher eigenfunctions display significant
differences which affect the quality of the fits to data. 
Remarkably, these differences diminish for higher eigenfunctions 
and for $n=41$ the two eigenfunctions almost overlap in this region. The reason
for this can be seen from Fig. \ref{nufig}, which shows that for the 
relatively low transverse momenta the differences in the frequencies between the
two models decreases with increasing eigenvalue number, so that if the 
infrared phases are  equal, the functions will be almost identical in this region.

\section{Results}
One of the main results of the previous paper \cite{KLRW} was  that we found  a simple power dependence between the infrared phase $\eta_n(k_0)$  and the eigenfunction number $n$. In this paper we use the same functional dependence,  defined for $\eta_n(\tilde{k}_0)$, where $\tilde{k}_0$ denotes an infrared scale at which the phase of the leading eigenfunction vanishes. The relation between the $\eta$ phase at  $\tilde{k}_0$ and $k_0$ is given by  eq.(\ref{eq:phrel}), where  the value of  $\tilde{k}_0$ should be close to $\Lambda_{QCD}$.  
Thus we take the parameterization
 \beq \eta_n(\tilde{k}_0) \ = \ \eta_0 \left(
 \frac{(n-1)}{(n_{max}-1)}\right)^\kappa,   
  \label{etan} \eeq
where $n_{max}$ is the number of eigenfunctions we use for the fit and
$\eta_0$ represents the total range (in units of $\pi$)
of infrared phases that are used for the fit. The value of the parameters $\tilde{k}_0$, $\kappa$ and $\eta_0$ are determined in the fit.

As explained in \cite{KLRW}, since the eigenvalue tends to zero
for large $n$, the  form of the phase given by eq.(\ref{etan}) means that
as a function of eigenvalue, $\omega$, the phase has a cut singularity
at $\omega=0$, i.e.
\beq
 \eta(\omega) \ =  \ \left(\frac{a}{\omega}\right)^\kappa+b +c\omega +d\omega^2 + \cdots
\label{etaomegapp}  \eeq
This allows the generalization of eq.(\ref{etan}) by treating all the  constants $a,b,c,d,\cdots$ in eq.(\ref{etaomegapp}) as free parameters. 
We have tested these parameterizations but find no
improvement in the quality of the fit although we introduced 
more parameters;  we therefore  used the simple version of the phase condition,  eq.(\ref{etan}), in all of our fits.  In ref. \cite{KLRW} we  fixed the value of the parameter $\eta_0$, which represents a total range of the $\eta$ variation. In the present evaluation, since we have to perform many more fits, we prefer to treat it as a free parameter, to assure a bias free evaluation of all cases.    Therefore we use in the fits the 3 parameters of eq.(\ref{etan}) and the 2 parameters from the  proton impact factor. For the impact factor we take the parameterization
\beq \Phi_p(k) \ = \ A k^2 e^{-b k^2}, \eeq 
 as in ref. \cite{KLRW}.
The fits were performed using the HERA data \cite{H1ZEUS} with $x \, < \, 0.01$
and $Q^2 \, > 8 \  \mathrm{GeV}^2$ or $Q^2 \, > 4 \  \mathrm{GeV}^2$. 

As  in \cite{KLRW} we find that 
there is an significant improvement of $\chi^2/N_{df}$ in the  $Q^2 \, > 8 \  \mathrm{GeV}^2$ region due to various higher order effects, such as the NLO
contribution to the photon impact factor and possibly also the proximity
of the saturation region. In this region we have a total of 108 data points and a total
of 5 parameters - so the number of degrees of freedom is $N_{df}= 103$. We therefore consider the $Q^2 \, > 8 \  \mathrm{GeV}^2$  region as our main investigation region and use the 
$Q^2 \, > 4 \  \mathrm{GeV}^2$ as a cross check.

    We investigated the fit quality as a function of the maximal number of eigenfunctions, $n_{max}$.  In contrast to the result of the analysis described in
\cite{KLRW}, we found that in the supersymmetric analysis  the best fit can be obtained with a somewhat smaller number of eigenfunctions;   only 100 (rather than 120) eigenfunctions are required to obtain the best  $\chi^2$. For the Standard Model the best fit is obtained with  $n_{max}=120$, but only  with a small difference in the  fit quality, $\chi^2=122.5(100)$ and $\chi^2=120.1(120)$.

\begin{table}
\begin{center}
\begin{tabular}{||c|c|c|c|c|c|c||}  \hline 
\begin{tabular}{c} SUSY Scale \\ (TeV) \end{tabular}
 & $\chi^2$ & $\kappa$ & $\tilde{k}_0 \ (GeV)$ & $\eta_0$ & A & b \\ \hline \hline
 3 & 125.7 & 0.555 & 0.288 & -0.87 & 201.2 & 10.6  \\  \hline
 6 &      114.1 &    0.575  &   0.279 &   -0.880 &     464.8 &      15.0 \\ \hline
 10 &   109.9 &   0.565 &  0.275 & -0.860 &   720.1 &   17.7  \\ \hline
 15 &   110.1 &  0.555 &  0.279 & -0.860 &   882.2 &  18.6 \\ \hline
 30 &  117.8 & 0.582 &  0.278 & -0.870 &  561.6 &   16.2 \\ \hline
 50 &  114.9 & 0.580 &  0.279 &  -0.870&  627.4 &   16.8 \\ \hline
 90 & 114.8  &  0.580  & 0.279 &  -0.870  &  700.2 &   17.5 \\ \hline
 $\infty$   &      122.5  &   0.600 &  0.274 &  -0.800  &   813.1  &   17.5 \\
\hline \hline
\end{tabular}
\caption{Fits for N=1 SUSY at different scales. The bottom row corresponds to
the Standard Model. All fits are performed with $n_{max}=100$. } \label{table8}
\end{center}
\end{table}

In Table \ref{table8} we show our fits for various SUSY thresholds as well
as the Standard Model. Let us first note that  the  $\tilde{k}_0$ values obtained in the unbiased fit, $\tilde{k}_0\sim 275$ MeV, are close to  $\Lambda_{QCD}$. 
At the same time the  value of $b$ implies that the proton impact factor peaks around $\Lambda_{QCD}$, as 
it should be in the self consistent description. This together with the relatively low $\chi^2$'s of all fits confirms the success of  our construction of the infrared boundary.



The quality of the fits shows a clear preference of the evaluation with SUSY effects; the fit for the Standard model
is  worse than the fits with SUSY thresholds larger than 3 TeV.
A SUSY threshold of 3 TeV, which is close to the reach of
LHC also gives  a worse fit. On the other hand for a SUSY threshold in the
region of 10 - 15 TeV, the quality of the fit is the best, but that for significantly larger SUSY thresholds
the fit quality worsens again.

Although the overal quality of the fit for all data with $Q^2 \, > \, 4   \, \mathrm{GeV}^2$
is significantly worse, for reasons outlined above,
the preference for  N=1 SUSY with the threshold  region of 10-15 TeV 
is also seen from a fit  to
 the $Q^2 \, > 4 \  \mathrm{GeV}^2$ data. In this $Q^2$ region there are 128 points and the $\chi^2$ 's of the best fits are 184.3 (3TeV),  164.5 (6TeV),   155.6 (10TeV), 152.6 (15TeV), 169.7 (30TeV), 164.7 (50TeV), 164.3 (90TeV). The best $\chi^2$ for Standard Model is 169.7.  The values of the fit parameters are similar to the values shown in Table \ref{table8}, for the $Q^2 \, > 8 \  \mathrm{GeV}^2$ region. 

 
The consistency of the fit results and a clear $\chi^2$ preference of the SUSY fits
 (with a  scale above 3 TeV) over that for the Standard Model  indicates that  supersymmetry
 improves the  data description and  
 suggests that some new physics similar to $N=1$ SUSY is present
in the 10 - 15 TeV region.

\section{Infrared Boundary: $\eta-\omega$ relation}
\begin{figure}
\centerline{\epsfig{file=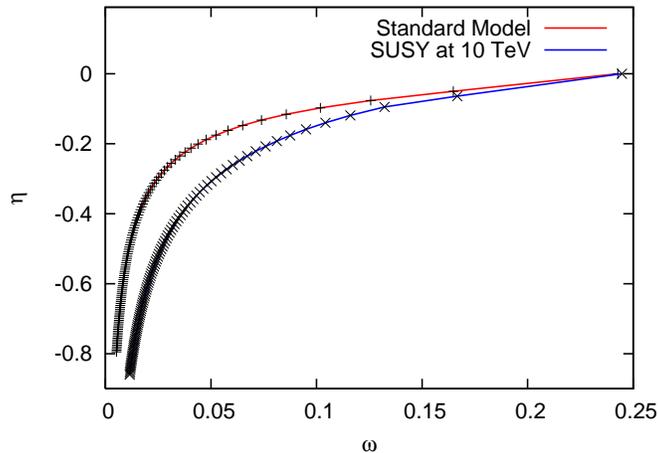, angle=270, width = 9 cm}}
\caption{ The eigenvalues and infrared phases for the Standard Model and  N=1 SUSY at 10 TeV, as determined at $k=\tilde{k}_0$. \label{etaomega} }
\end{figure}

The infrared boundary condition that leads to a discrete spectrum
can be expressed as the ansatz that the phase at some infrared transverse momentum,
$\tilde{k}_0$, is a fixed function, $\eta(\omega)$ of the eigenvalue $\omega$, of the form given by eq.(\ref{etaomegapp}), imposed by the infrared properties of QCD.
The discrete pomeron spectrum is driven by this function. Therefore, in Fig. \ref{etaomega} we show
the values of the eigenvalues $\omega_n$ and the infrared phases $\eta_n$ (in units of $\pi$)  both for
 the Standard Model and $N=1$ SUSY at a threshold of 10 TeV.
 The numerical values of the parameters
on the RHS of eq.(\ref{etaomegapp})
turn out to be substantially different in the two cases. This function 
constitutes the infrared boundary conditions on the eigenvalues of the BFKL kernel. 
As explained above the eigenvalues are somewhat larger for the SUSY model, but both of these
functions have a cut  at $\omega=0$, the order of the cut singularity being a little less
in the case of the SUSY model (the dip is not so steep).

As was discussed in ref.\cite{KLRW} the appearance of the singularity in the $\eta-\omega$ relation indicates that some important contribution to the perturbative expansion at very large transverse momenta is missing. Therefore it is interesting to observe that the introduction of SUSY softens somewhat the observed singularity ($\kappa$ is reduced from 0.6 to 0.56) and at the same time reduces the number of eigenfunctions required
 - it is a step towards a description of data using only few discrete pomerons. 
This could also indicate  that  there exist other, even stronger symmetries at very high energies, which are missing in the present evaluation, and which are responsible for the remaining singularity of the phase $\eta(\omega)$.

\section{Summary}
 In our previous paper \cite{KLRW} we have shown that DPS gives a very good description of the low-$x$ HERA data and 
it was suggested that this fit may have sensitivity to  BSM physics.  
This proposed sensitivity emerged from the fact that the higher eigenfunctions have support over
a very large range (see eq. (\ref{range})), extending from the infrared region to way above the threshold for
any new physics and that through the required phase-matching process, the low energy
behaviour of these eigenfunctions depends on their high-energy behaviour.

In this paper, we have shown that this is indeed the case. The introduction of N=1 SUSY at some threshold alters the value of the $\beta$-function and hence the rate of the running of the coupling. Furthermore there are contributions to the
NLO characteristic function of the BFKL equation from the   SUSY partners. Since the properties of the discrete pomeron are
determined from a combination of the running coupling and the characteristic function, the eigenfunctions of the  BFKL kernel are significantly affected by the introduction of SUSY. Notwithstanding the fact that the proposed  SUSY scale is considerably above the scale probed at HERA, the altered
high-energy behaviour of the eigenfunctions feeds into the low-energy, as well as generating
a somewhat different spectrum of eigenvalues. 

The discrete spectrum depends on the treatment of the infrared boundary condition, which is now more involved due to the fact that we are using the two-loop $\alpha_s$ running, instead of one-loop, as in \cite{KLRW}. Constructing this boundary we took the most conservative approach of using  perturbative QCD as a guideline at every step.
Our previous paper  was devoted to the task of finding the  relation between the eigenfunction number and the phase of the eigenfunction oscillations which is essentially of the non-perturbative origin.
Notwithstanding the substantial differences between the eigenfunctions with and without SUSY,
we find here that the best fit is obtained
using the same form of this dependence as was used in that paper, although other forms for this 
dependence were attempted without improving the fit quality.

Together, the different spectrum of eigenvalues and the different  shapes of the eigenfunctions
 in turn affect the parameters
of the fit to data and also the quality of the fit.  
The main result of this paper is that if the
SUSY threshold is introduced at 10 -15 TeV,
the fit was significantly  better than that of the Standard Model ($\chi^2=110$ as compared to $\chi^2=122$ for 108 points and 103 degrees of freedom, i.e $\chi^2/N_{df} = 1.06 $ vs 1.18). 
On the other hand, the  introduction of SUSY at the threshold of 3 TeV, just within the reach of LHC,
generates a fit which is no better than the fit obtained from the Standard Model.

It is pertinent to emphasize the qualitative difference between the fit
obtained here and the usual DGLAP fit \cite{DGLAP}. 
Over the low-$x$ region, this DGLAP fit obtains a slightly better value of $\chi^2$
per degree of freedom ($\chi^2/N_{df} \, \sim \, 0.95$).
However, the DGLAP parameterization is designed to cover
the entire range of $x$, whereas ours is only valid for sufficiently low-$x$
where an expansion in $\ln(1/ x)$ is valid. The improved quality of the fit of \cite{DGLAP}
is likely to be due to the terms with positive powers of $x$ that are present in that fit.
The important qualitative difference between the two fitting procedures is that
the parameters obtained in the DGLAP  approach are unaffected by any new physics 
at high-energy thresholds - their prediction for the structure functions would remain
unchanged until the threshold (in $Q^2$) for new physics were reached. On the other hand, as we
have emphasized in this paper, the values of our parameters are affected by new physics
thresholds and consequently the $Q^2$ evolution above the fit region will always be affected
by such new physics. 
This considerably stronger predictive power of the  BFKL equation is not only due to the fact that it is  a different type of  evolution equation, but also that it describes the dynamics of the gluon-gluon interaction which (after accounting for the infrared boundary conditions) produces the two-gluon quasi-bound states with a non trivial spectrum of singularities in the $j$-plane.

The method described in this paper opens a new possibility to use high precision experiment
 to search for new physics  at energy scales considerably larger than the scales at which the experiments
  are performed. It can be applied to any low-$x$ process which was measured with comparable accuracy to the HERA $F_2$ data, like the Drell-Yan or W and Z production at LHC.  The application of this method to LHC data could lead to higher sensitivity due to substantially higher scales involved. The discrete pomeron solution provides a unique tool for such an investigation owing to the fact that the construction of the eigenfunctions is based on the quantum mechanical approach in which  the extremely high energy (up to the Planck scale)
 and low energy behaviour of its wave functions are  intimately connected.  
\bigskip

{\noindent}{\bf Acknowledgements:} \\
The authors are grateful to the Marie Curie Foundation for an IRSES grant,  LOWXGLUE Project 22498,
which has facilitated this collaboration. Two of us (HK and DAR) wish to thank
the St. Petersburg Nuclear Physics Institute for its hospitality while this work
was carried out.

\end{document}